\documentclass[aps,prd,eqsecnum,twocolumn,twoside,showpacs]{revtex4}
\usepackage{amsfonts}
\begin{document}

\title{
A search for the minimal unified field theory.
I. Parallel transport of Dirac field.  }
\author{ Alexander Makhlin}
\email[]{amakhlin@comcast.net}
\affiliation{ Rapid Research Co., Southfield, MI, USA}
\date{September 9, 2004}

\begin{abstract}

 Two real vector fields are revealed as spin connections of the spinor
 field, which is  introduced as a representation of the local Lorentz
 group  by Dirac  spinors. One of these fields is identified  as the
 Maxwell field. Another one is  the {\em axial field},
 which is  sufficient  to describe parity nonconservation effects in
 atoms and it is  sensitive to the second internal degree of freedom
  of the  Dirac spinor field. The polarization associated
 with this degree of freedom  is a distinctive characteristic of the
 Dirac spinor field in spatially closed  localized states. The
 short-distance behavior of the axial field can lead to a  Dirac
 Hamiltonian that is not self-adjoint and trigger a
 ``falling onto a centre'' phenomenon. In the second part of this
 work, the axial field will be linked to the gravity.
\end{abstract}

\maketitle

\section{ Introduction\label{sec:Sec1}}

In the late 1920's, V. Fock formulated a vigilant approach
\cite{Fock1} to the problem of ``minimal interaction'' between the
spinor matter and the Maxwell field, which was not based on
Weyl's {\em ad hoc} declaration of gauge invariance \cite{Weyl}. Fock
derived the Maxwell field as a part of the spin connection of the Dirac
field.  The more confidence we gain that spinor fields should be
thought of as the fundamental form of matter, the more rationale
we have to explore an immense potential of a so consistent, economic
and elegant method. There is  no doubt that the spinor fields are the
most important objects in the physics of stable matter (the Pauli
principle is decisive for its actual existence). Moreover, the
two-component spinors represent the light cone of special relativity
in a most straightforward way, for their currents are light-like. As a
result, the dynamics of spinors dominate those physical phenomena where
sharp space-time localization occurs.

In this paper I show that, following the Fock method, one can derive
the existence of a second real vector field, which is minimally
coupled to the axial vector current of the Dirac spinor field. From a
theoretical perspective, this field seems to be just overlooked, and
(in 1920's) there was no experimental evidence of its
existence. For the lack of an established term, I shall refer to it
as an {\it axial field}. This field is not gradient invariant and thus
should be massive. At the first glance, it very much resembles the
neutral vector field of the electro-weak theory. The axial field
resolves the second internal degree of freedom of the Dirac spinor
field, the existence of which was pointed out by Fock \cite{Fock3}. The
presence of the axial field in the Dirac equation reasonably
describes the (well known by now) parity non-conservation (PNC)
phenomena in atomic physics. The most distinctive physical feature of
this axial field is that its Lorentz force pulls the left and right
components of a 4-spinor in opposite directions. This causes an
additional polarization of the Dirac field and  additional types of
spinor modes (with quantum numbers that correspond to compact
objects) become possible. Like the electromagnetic field, the axial
field emerges as one of the ingredients of parallel transport of
Dirac spinors; it is a kinematic effect stemming from the complex
nature of the spinor field and a large number of its polarization degrees
of freedom. All immediately anticipated physical effects of the axial
field seem to be known or expected. Surprising is the fact that this
field is derived at the most basic level of Lorentz invariance and
that it was not motivated by any phenomenological input, e.g., a
symmetry observed in particular processes.

A major focus of this first paper is on the derivation and classical
aspects of the axial field. This includes a detailed analysis of
parallel transport of the Dirac spinors, simultaneous derivation of
the Maxwell- and axial- fields as the components of the spin
connections, and an analysis of the effect of the {\it external}
axial field on the spherically-symmetric states (hydrogen-like atom).
The fact that the axial field makes the Dirac equation extremely
singular is left aside in this paper. This fact is merely
demonstrated, which then leads to more subtle aspects of the axial
field, including its role in auto-localization of the Dirac field and
its relation to gravity, which are discussed in the second paper
\cite{paper2}.

\section{ Parallel transport of the Dirac field.\label{sec:Sec2}}

\renewcommand{\theequation}{2.\arabic{equation}}
\setcounter{equation}{0}

To define the derivative of a vector or tensor field one has
to simultaneously compare many components at two neighboring points.
Only for vectors given in Cartesian coordinates does the notion of
parallel transport have no ambiguity. One has to account for a
continuous rotation of the local coordinate hedgehogs along a
displacement path and include a connection into the covariant
derivative of a vector even if the same flat space-time is
parameterized by oblique coordinates . The covariant derivative of
a vector can then be derived in a relatively simple way because
rotation of a vector at a given point follows the rotation of the
{\em local} coordinate axes. There is no equally simple rule for
spinors.

\subsection{ Tetrad formalism}

The tetrad formalism \cite{Eisenhart} directly inherits the principle
of local equivalence of the theory of relativity. The four
components $V_\mu$ of a usual (coordinate) vector are replaced by four
coordinate scalars $V_a$, which behave as the vectors with respect to
Lorentz rotations $\Lambda_a^b(x)$ of {\it local} coordinate axes,
$V_\mu=e^a_\mu V_a,~~~V_a(x)\to\Lambda_a^b(x) V_b(x)~$. The four tetrad
vectors $e^a_\mu$ are the Lorentz vectors and coordinate vectors
at the same time. The curvilinear metric ${\rm g}_{\mu\nu}(x)$ is
connected with the local metric $g_{ab}$ of Minkowski space by
\begin{eqnarray}\label{eq:E2.2}
{\rm g}_{\mu\nu}(x)=g_{ab}e^{a}_{~\mu}(x)e^{b}_{~\nu}(x),~~~~~~~~~\\
g^{ab}= {\rm g}^{\mu\nu}(x) e^{a}_{~\mu}(x)e^{b}_{~\nu}(x)=
{\rm diag}[1,-1,-1,-1]~,\nonumber
\end{eqnarray}
and the latter is the same at all space-time points. The coordinate
and the tetrad components of the covariant derivatives are
$\nabla_\lambda V_\mu= \partial_\lambda V_\mu
-\Gamma^\nu_{\mu\lambda} V_\nu$,
and
$D_a V_b=e^\lambda_a (\nabla_\lambda V_\mu)e^\mu_b= \partial_a V_b -
\omega_{bca} V^c$,
respectively. In these equations, $\Gamma^\nu_{\mu\lambda}$ are the
Christoffel symbols, $\partial_a=e_a^\mu\partial_\mu$ is the derivative
in direction $a$, and $\omega_{abc}$ are the Ricci rotation coefficients,
$\omega_{bca}=(\nabla_\mu e_{b}^{\nu})e_{c\nu}e_{a}^{\mu}=-\omega_{cba}~.$

It is useful to keep in mind that the absolute differential,
$DV_a$, of a vector $V_a$ is the principal linear part of the vector
increment with respect to its change in the course of parallel
transport along the same infinitesimal path. (A bonus for
this definition is that $DV_a$ is also a vector.) Therefore,
the parallel transport just means that $DV_a=0$, and the same definition
should be adopted for spinors as long as they can represent vector
observables. As a matter of fact, the tetrad representation most
adequately reflects the local nature of vector fields in relativistic
field theory. The Lorentz transformation and the definition of
parallel transport for spinors are far less obvious, because their
components are not directly connected with the vectors of the coordinate
axes. Therefore, spinors should always be treated as coordinate scalars.

All observables associated with the spinor fields are Lorentz tensors,
which are bilinear forms built with the aid of Dirac matrices.
Within the tetrad formalism, these matrices should be treated as
coordinate scalars: {\em They are the same at each local Lorentz frame
at each world point.} We use an old-fashioned convention that corresponds
to the spinor representation. The conjugated spinor is  $\psi^+$.
The sixteen Dirac matrices  can be conveniently arranged as the products
like $\rho_a \sigma_b$, $~~a,b=0,1,2,3$. The basic matrices $~\rho_i~$
and $~\sigma_i$, $~i=1,2,3$, were introduced by Dirac \cite{Dirac1};
we reserve various notation, $~\rho_{0}=\sigma_0={\bf 1}~$,
for the same unit matrix. The other notation for the Dirac
matrices are: $\alpha^a =(\alpha^0,\alpha^i)$(with
$\alpha^0=1$, $\alpha_i=\gamma^0\gamma^i=\rho_{3} \sigma_i$),
$\rho_1=\beta=\gamma^0$, $\rho_2=-i\gamma^0\gamma^5$,
$\rho_3=-\gamma^5$. The $4\times 4$ matrices $\sigma$ and $\rho$
satisfy the same commutation relations as the Pauli matrices, and
all matrices $\sigma$ commute with all matrices $\rho~$:
$ \sigma_i\sigma_k=\delta_{ik}+i\epsilon_{ikl}\sigma_k$,
$\rho_a\rho_b=\delta_{ab}+i\epsilon_{abc}\rho_c~$,
$\sigma_i\rho_a -\rho_a\sigma_i =0$.

Three $4\times 4$ matrices $\sigma_i$ have a block form with the standard
$2\times 2$ Pauli matrices $\tau_i$,
\begin{eqnarray}
\tau_1=\left( \begin{array}{c c} 0  & 1  \\ 1& 0 \end{array} \right),
~ \tau_2=\left( \begin{array}{c c} 0 &  -i  \\
i & 0 \end{array} \right),~
 \tau_3=\left( \begin{array}{c c} 1  &  0  \\ 0& -1 \end{array} \right),
\label{eq:E2.7}\end{eqnarray}
along the block-diagonal, i.e. $\sigma_i={\bf 1}\otimes\tau_i$. Matrices
$\rho$ are the same products in inverse order,
$\rho_a=\tau_a\otimes{\bf 1}$,
\begin{eqnarray}
\sigma_i=\left( \begin{array}{c c} \tau_i  &  0 \\
0& \tau_i\end{array}\right),~~
\rho_1=\left( \begin{array}{c c} 0  & {\bf 1}  \\
 {\bf 1}& 0 \end{array}\right),\nonumber\\
\rho_2=\left( \begin{array}{c c} 0 &  -i\cdot{\bf 1}  \\
 i\cdot{\bf 1} & 0 \end{array} \right),~~
 \rho_3=\left( \begin{array}{c c} {\bf 1}  &  0  \\
  0& -{\bf 1} \end{array} \right).
\label{eq:E2.8}\end{eqnarray}

\subsection{ Parallel transport of Dirac spinors and spin
connections}

One can look at the spinors as the {\em fields} and differentiate them
only after having stated the rules for parallel transport. To derive these
rules we must rely on various tensors, which can be arranged as a matrix,
$\psi^+ \rho_a \alpha^b \psi$.   It is most {\em natural} to require
the vector current,
\begin{eqnarray}
j_a =\psi^+ \alpha_a \psi \equiv  \psi^+ (1,\rho_3\sigma_i)\psi
\equiv {\bar \psi} \gamma_a \psi~,
 \label{eq:E2.11}\end{eqnarray}
to be a Lorentz vector. Indeed, its temporal component is
the unit quantum mechanical operator, which
thus commutes with all other operators, including the Hamiltonian,
and by all means, represents a unique conserved quantity. This
operator is allied with the time-like tetrad vector that designates
the direction of Hamiltonian evolution and thus defines all observables
and, ultimately, the physical content of a theory. The vector $j_a(x)$
is transformed as $j_a(x) \to \Lambda_{a}^{~b}(x)j_b(x)$ under
the local Lorentz rotation, and its variation under the displacement
$dx^\mu$ is $~\delta j_\mu=\Gamma^\nu_{\mu\lambda}j_\nu dx^\lambda$.
The tetrad components of this vector change by
\begin{eqnarray}
\delta j_a= \omega_{acb} j^c ds^b= \omega_{acb} \psi^+ \alpha^c \psi  ds^b~,
 \label{eq:E2.12}\end{eqnarray}
when this vector is transported at $ds^a$.
Let matrix $\Gamma_a$ (the spin connection) define the change of the
spinor components in the course of the same infinitesimal displacement,
$\delta \psi= \Gamma_a \psi ds^a$, $~\delta \psi^+=\psi^+ \Gamma^+_a ds^a$.
This gives yet another expression for $\delta j_a$,
\begin{eqnarray}
\delta j_a= \psi^+ (\Gamma^+_b \alpha_a+\alpha_a\Gamma_b)\psi ds^b~.
\label{eq:E2.14}\end{eqnarray}
This must be the same as Eq.~(\ref{eq:E2.12}). Hence,
the equation that defines $\Gamma_a$ is
\begin{eqnarray}
\Gamma^+_b \alpha_a+\alpha_a\Gamma_b= \omega_{acb} \alpha^c ~.
\label{eq:E2.15}\end{eqnarray}
The matrices $\{1,\rho_a,\sigma_i,\rho_a \sigma_i \}$ form a complete set,
and all of them, except for the unit matrix, are traceless; the square of
each of these matrices is a unit matrix. Therefore, the spin connection
can be expanded as
\begin{eqnarray}
\Gamma_l= a_l +\sum_a r^l_a \rho_a +\sum_j z^l_j \sigma_j
+\sum_{a,j}f^l_{a,j} \rho_a\sigma_j .
\label{eq:E2.17}\end{eqnarray}
This results in a system of equations for the coefficients that has the
following solution,
\begin{eqnarray}
a_b= -ieA_b,~~~r_{3~b}=-i g \aleph_b,~~~\nonumber\\
 f^b_{3,j}=-{1\over 2}\omega_{0ib},~~~
z^b_k=-{i\over 4}\epsilon_{imk}\omega_{imb},
\label{eq:E2.18}\end{eqnarray}
where $A_b$ and $\aleph_b$ are the tetrad components of two real
vector fields. The origin of these fields is very simple:
Eqs.~(\ref{eq:E2.15})
yield  only two relations, ${\rm Re} [a^b]=0$ and ${\rm Re} [r_3^b]=0$,
for the complex coefficients $a^b$ and $r_3^b$, which leave their
imaginary parts arbitrary functions. Thus,
\begin{widetext}
\begin{eqnarray}
\Gamma_b(x)=-ieA_b(x) -ig\rho_3 \aleph_b(x)
-{1\over 2}\omega_{0ib}(x)\rho_3\sigma_i
-{i\over 4}\epsilon_{0kim}\omega_{imb}(x)\sigma_k~,
\label{eq:E2.19}\end{eqnarray}
where the last two terms can be combined into
$\Omega_b(x)=(1/4)\omega_{cdb}(x)\rho_1\alpha^c\rho_1\alpha^d~$.
It is instructive to see the explicit matrix form of the
spin connection ,
$$\Gamma_b= \left( \begin{array}{c c}-ieA_b & 0  \\
0 & -ieA_b \end{array} \right)
+ \left( \begin{array}{c c}-ig\aleph_b  & 0  \\
0 & ig\aleph_b \end{array} \right)
-{1\over 2}\omega_{0ib}
\left( \begin{array}{c c} \tau_i  & 0  \\ 0 & -\tau_i \end{array} \right)
-{i\over 4}\epsilon_{imk}\omega_{imb}
\left( \begin{array}{c c} \tau_k  & 0  \\ 0 & \tau_k \end{array}
\right)~.$$
\end{widetext}
The upper and the lower rows of this matrix are the spin connections
for the left (not dotted) spinor $\xi^\alpha$, and for the right
(dotted) spinor $\eta_{\dot\alpha}$, respectively.  The function
$A_a(x)$ is the same for both spinors; it is immediately recognized
as (and will be proved to be) the vector potential of the
gradient-invariant Maxwell field.

The signs in the second and third terms alternate between the left
and right spinors. One can easily understand the difference in the
third term by recalling that $\omega_{abc}ds^c$ is the angle at which
the local tetrad is rotated in the $ab$ plane when it is transported
by infinitesimal $ds^c$. The rotation $\omega_{0ic}$ is
the Lorentz boost; it {\em must} have opposite signs for the left and
right spinors.  The spin connection $\rho_3\aleph_a(x)$ also
differentiates between the left and right spinors  and it will become
the field that interacts with the axial current. This is the most
general form of a spin connection allowed by Lorentz invariance,
and all its components are equally important. In order for the Dirac
field to form compact objects, this general form must be restricted.
This will be done in the next paper \cite{paper2}.

Next, we have to find the transformation properties of the twelve
remaining elements of the type $\psi^+ \rho_a\sigma_b\psi$.  The
array $$~{\cal J}_a=\psi^+ \rho_3\alpha_a \psi \equiv  \psi^+
(\rho_3,\sigma_i)\psi \equiv {\bar \psi}\gamma^5 \gamma_a \psi~$$ is
known as the axial current. Using $\Gamma_a$ of Eq.~(\ref{eq:E2.19})
we find that $\delta {\cal J}^{}_b=\omega_{bca}{\cal J}^c ds^a $.
Hence, the axial current $~{\cal J}^{}_b~$ is indeed transported as a
Lorentz (pseudo-)vector. The two other densities, $~{\cal S}=
{\bar \psi}\psi \equiv \psi^+ \rho_1 \psi$ ~and
$~{\cal P}={\bar \psi}\gamma^5\psi \equiv \psi^+ \rho_2 \psi$,
are expected to behave as Lorentz scalars. This is indeed the case in
the sense that their parallel transport does not depend on the
parameters of the Ricci rotations,
$$ \delta {\cal S} = -2g {\cal P} \aleph_a ds^a, ~~{\rm and}
 ~~\delta {\cal P} = 2g {\cal S} \aleph_a ds^a .$$
However, in the course of parallel transport the axial
field $\aleph$ mixes them, i.e.,
\begin{eqnarray}
{D{\cal S}\over ds^a}={\partial{\cal S}\over \partial s^a}+
2g\aleph_a{\cal P},~~
{D{\cal P}\over ds^a}={\partial{\cal P}\over \partial s^a}-
2g\aleph_a{\cal S}.
\label{eq:E2.22a}\end{eqnarray}
(Hereafter, this work substantially deviates from Fock's
paper \cite{Fock1}. Fock requires $\delta {\cal S}=0$ and
$\delta {\cal P}=0$ though there is no observables that
support these conditions.)
The skew-symmetric Lorentz tensor $M_{ab}$ is supposed to be transported as
$$\delta M_{ab}=\omega_{acd}M^c_{~b} ds^d +
\omega_{bcd}M_a^{~c} ds^d,$$
and an explicit calculation shows that the increment of this tensor
with the components
$$M_{0i}= -i\psi^+ \rho_2\sigma_i \psi
\equiv {\bar \psi}\gamma^0\gamma^i\psi~$$ and
$$M_{ik}= -i\epsilon_{ikm}\psi^+ \rho_1\sigma_m \psi
\equiv-\psi^+ \rho_1\sigma_i\sigma_k \psi
 \equiv{\bar \psi}\gamma^i\gamma^k\psi,$$
has the correct Lorentz part. However, this increment
also acquires additional terms due to the field $\rho_3 \aleph_a$,
$$\delta M_{0i}=(\omega_{0jb}M_{jb}-
\omega_{imb}M_{0m}) ds^b + g \epsilon_{ijm}M_{jm}
\aleph_b ds^b,$$
$$\delta M_{ik}=\!(-\omega_{icb}M^c_{~k}+
\omega_{kcb}M_i^{~c}) ds^b\!
-\! 2 g \epsilon_{ikm}M_{0m} \aleph_b ds^b.$$
Therefore, contrary to naive expectations, the parallel transport of
scalar, pseudoscalar and skew-symmetric tensor results in  additional
terms, which are proportional to $\aleph_\mu=e_\mu^a\rho_3\aleph_a$.
This field does not affect parallel transports of the vector current
and axial current (antisymmetric tensor of rank three). However,
the axial field $\aleph$ does affect parallel transport of
the skew-symmetric tensor $M_{ab}$ of rank two and its scalar
invariants ${\cal S}$ and ${\cal P}$, all of which are built with
the aid of the off-diagonal matrices $\rho_1$ and $\rho_2$.
These matrices mix the left and right spinors in the equation of motion
and various observables. Mathematically, the two additional fields,
$A_a(x)$ and $\aleph_a(x)$, correspond to extra degrees of
freedom that are left to spinors by Lorentz invariance of the tensor
observables.

\subsection{ Parallel transport and the Lorentz transformations.}

The redundant invariance that has appeared in the law of parallel
transport must show up in Lorentz transformation of spinors also.
Indeed, let us  derive the law of the Lorentz transformation for a spinor,
$ \psi\to S\psi $, $~\psi^+\to\psi^+ S^+~$, starting from the same
current $j_a=\psi^+ \alpha_a \psi $, which is transformed as
 $j_a \to \Lambda_{a}^{~b}(x)j_b$. This gives
\begin{eqnarray}
 S^+\alpha_a S=\Lambda_{a}^{~b}(x)\alpha_b~,
\label{eq:E2.26}\end{eqnarray}
which is the equation that determines $S$. The matrix $S(x)$ can have one
of two forms,
\begin{eqnarray}
 S=\left(\begin{array}{c c}\lambda & 0\\0 &
(\lambda^+)^{-1}\end{array}\right),~~~{\rm or}~~~
 S=\left(\begin{array}{c c}\lambda & 0\\0 &
\lambda^*\end{array}\right); ~~\nonumber\\
\lambda=\left( \begin{array}{c c}\alpha  & \beta
\\ \gamma & \delta \end{array} \right),
~~~\alpha\delta-\beta\gamma=1~,~~~~~
\label{eq:E2.26a}\end{eqnarray}
The "dotted" two-component spinor $\psi^R=\eta$ must be transformed
either as dual to $\psi^L=\xi$ (i.e., by means of matrix
$(\lambda^+)^{-1}=\tau_2\lambda^*\tau_2$) or as the conjugated to
it (by means of matrix $\lambda^*$). Only then
will $(1/2,0)$ and $(0,1/2)$ be the two
different representations of the Lorentz group. This transformation
should preserve Eq.~(\ref{eq:E2.15}) in its coordinate form.
This form can be derived by introducing
$\alpha_\mu(x)=e_\mu^a \alpha_a$ and $\Gamma_\mu=e_\mu^a \Gamma_a$,
and accounting for the fact that the tetrad vectors are covariantly
constant,
$$~D_\mu e^b_\nu =\nabla_\mu e^b_\nu-e^a_\mu \omega^b_{~ca}e^c_\nu=0,
~~\nabla_\lambda e^b_\mu= \partial_\lambda e^b_\mu
-\Gamma^\nu_{\mu\lambda} e^b_\nu$$
(which is just another way to define the Ricci coefficients).
Hence, the matrices $\alpha_\mu(x)$ are covariantly constant also , i.e.
$D_\mu \alpha_\nu=D_\mu \alpha_b e^b_\nu=  \alpha_b D_\mu e^b_\nu
=0$, or, explicitly,
\begin{eqnarray}
D_\mu \alpha_\nu=\nabla_\mu \alpha_\nu +\Gamma^+_\mu \alpha_\nu
 + \alpha_\nu \Gamma_\mu =0~,
\label{eq:E2.31}\end{eqnarray}
which is just a translation of Eq.~(\ref{eq:E2.15}) from the tetrad basis
into the coordinate basis.

The covariance of (\ref{eq:E2.31}) implies that the Lorentz
transformation
of a spinor,
$\psi(x) \to\tilde{\psi}=S(x)\psi(x)$, must be accompanied by a similar
transformation of the the matrices $\alpha_\mu(x)$, i.e.
${\alpha}_\mu(x)\to\tilde{\alpha}_\mu(x)=S^{-1}(x)\alpha_\mu(x) S(x)$.
Hence, in order that Eq.~(\ref{eq:E2.31}) be the same in any local Lorentz
frame, the spin connections should transform as
\begin{eqnarray}
{\Gamma}_\mu\to \tilde{\Gamma}_\mu=S \Gamma_\mu S^{-1} +
(\partial_{\mu}S)S^{-1}~,
\label{eq:E2.32}\end{eqnarray}
where $S$ is the solution of Eq.~(\ref{eq:E2.26}).
The first term of this formula ``rotates" the spinor indices at each
point, while the second inhomogeneous term adds a gradient.

To compare $S$ with the law of parallel transport, consider
the infinitesimal Lorentz transformation, where
$\Lambda_{ab}=\delta_{ab}+\varpi_{ab}$, and $ S=1+T$.
Then Eq.~(\ref{eq:E2.26}) becomes
\begin{eqnarray}
T^+ \alpha_a+\alpha_a T =\varpi_{ab}\alpha_b~,
\label{eq:E2.28}\end{eqnarray}
which is exactly Eq.~(\ref{eq:E2.15}) with
$\omega_{abc}d s^c$ replaced by an infinitesimal
$\varpi_{ab}=-\varpi_{ba}$. The most general solution of this equation
that has the required form (\ref{eq:E2.26a}) is
\begin{eqnarray}
T=ie\alpha(x)-{1\over 2}\varpi_{0k}(x)\rho_3\sigma_k -{i\over
4}\epsilon_{0kim}\varpi_{im}(x)\sigma_k~.
\label{eq:E2.29}\end{eqnarray}
One can recognize the same pattern as
in Eq.~(\ref{eq:E2.19}), i.e. the left and right spinors are transformed
differently. Thus, the local Lorentz rotation of the spinor field acts in
agreement with the transformation of its spin connections. Note,
however, that Eq.~(\ref{eq:E2.29}) misses the counterpart of the term
$\rho_3\aleph_a(x)ds^a$ in the spin connection even though the
latter is minimally  coupled to the spinor field. This is a consequence
of the above requirement that the matrix $S$ has the
form of Eq.~(\ref{eq:E2.26a}).

\section{ Equations of motion and  conservation laws\label{sec:Sec3}}

\renewcommand{\theequation}{3.\arabic{equation}}
\setcounter{equation}{0}

It is common to begin field-theory calculations from a Lagrangian
that has an alleged symmetry of the dynamical system under
investigation and to derive the equations of motion and conservation
laws using a variation principle. Because of a new axial field in the
covariant derivative of the Dirac spinors we have to proceed more
gradually. Namely, we must establish the equations of motion as the
first step, then to check their compliance to the data and derive
the identities that have a form of conservation laws, and only after
that to write a Lagrangian that reproduces the same results.

\subsection{ Equation of motion (Dirac equation revisited).}

In view of a new spin connection, the equation of motion
of the Dirac field has to be revisited. In fact,
we have nothing else at our disposal that can be used
to create an equation of motion, except for the covariant derivative
of the spinor field. In a tetrad basis it was found to be
$D \psi= D_a \psi d s^a = (\partial_a \psi -\Gamma_a \psi) d s^a~,$
with the connection given by Eq.~(\ref{eq:E2.19}). The structure of
this spin connection is two-fold; its spin indices are being used to
parameterize rotations of the local tetrad basis by means of Pauli
matrices. Its Lorentz index shows the direction of parallel transport.
The first step is to parameterize the Lorentz index $a$ by a spinor
and thus to convert this derivative entirely into a spinor
representation. This should be done exactly
as for the covariant coordinate vector $x_a$. Actually, we have to use
$ x_a \to x_{\alpha{\dot\beta}}={\bf 1}x_0+{\vec\tau}\cdot{\vec x}$
 and
 $~x_a \to x_{{\dot\alpha}\beta}={\bf 1}x_0-{\vec\tau}\cdot{\vec x}$
for the left and  right  spinors, respectively. This simultaneous
conversion for two components of the Dirac spinor is carried out
by means of matrix
$~\alpha_a=(1,\rho_3\sigma_i)$ -- the left and right spinors are Lorentz
transformed {\em differently}.  An obvious conjecture is that the
two spinors {\em do not evolve independently}. Otherwise an electron
would be composed of two dynamically disconnected fields each having
its own light-like current. Thus, the equation of motion for the spinor field
must be compiled according to the correspondence principle, which can take
different forms. The Dirac form that appeals to the relativistic
$p^2=E^2-\vec{p}^2=m^2$ can be found  in almost any textbook. Another
form can be traced back to the Fock proper-time argument \cite{Fock2}.
It employs virtually the same classical limit but in dimensionless form,
$u^2=u_0^2-\vec{u}^2=1$. (On can even view it as a prototype of the
commutation relations, $\alpha^a\rho_1\alpha^b+\alpha^b\rho_1\alpha^a
=2\rho_1g^{ab}$, that sets the Clifford algebra of the Dirac matrices.)
It is equivalent to the previous one as long as
$P^\mu=m u^\mu$, but it does not allow for the continuous limit of
$m^2\to 0$. This second form is more restrictive -- it is a requirement
that the electron has a rest frame, and a continuous limit of $u^2\to 0$
just does not exist. Assuming this form, and taking the linear relation,
$u_\mu P^\mu=m$ as a prototype, we can write
the following version of the Dirac equation,
\begin{eqnarray}
\alpha^a(\partial_a \psi -\Gamma_a \psi)+im\rho_1 \psi=0~,
\label{eq:E3.2}\end{eqnarray}
with the spin connection again given by Eq.~(\ref{eq:E2.19}). This
equation includes an
additional vector field $\aleph_a(x)$. The existence of such a field
must be confirmed by data at the same level of confidence as in the
case of the Maxwell field. It turns out that the axial
field is responsible for a dynamic auto-localization of the Dirac field,
so that the new equation remains very close to its classical
prototype. The conjugated equation reads as
\begin{eqnarray}
(\partial_a \psi^+ - \psi^+ \Gamma_a^+ )\alpha^a- i m \psi^+ \rho_1=0~.
\label{eq:E3.3}\end{eqnarray}
The differential operators of these equations are Hermitian
(symmetric), which is confirmed by the conservation of the vector
current $j^\mu$. The question if they can be self-adjoint in the
presence of an axial field is not trivial and will be addressed later on.

\subsection{ Conservation identities for vector currents.}

The equations of motion (\ref{eq:E3.2}) and (\ref{eq:E3.3})
immediately lead to two well-known identities,

\begin{eqnarray}
\nabla_\mu j^\mu={1\over\sqrt{-{\rm g}} }
\partial_\mu[\sqrt{-{\rm g}}\psi^+ \alpha^\mu \psi]=0,
\label{eq:E3.4}\end{eqnarray}
and
\begin{eqnarray}
\nabla_\mu {\cal J}^\mu = {1\over\sqrt{-{\rm g}} }
\partial_\mu[\sqrt{-{\rm g}}\psi^+ \rho_3\alpha^\mu \psi]=
2m{\cal P}.
\label{eq:E3.5}\end{eqnarray}
The first of them, indicates the conservation of the probability
current.
The second one states that the axial current cannot be conserved,
in a full compliance with its space-like nature. A major factor
here is a finite pseudoscalar
density (the mass $m$ is no more than a scale parameter),
which is a measure of a dynamic interplay between the left
and right components of the Dirac spinor. This factor
become active when a Dirac particle experiences an
acceleration and its field  deviates from a perfect parity-even
configuration encoded in the conventional Dirac equation
(Cf. \cite{Berest,Ryder}).

\subsection{ Energy-momentum conservation.}

Physically, the energy momentum tensor is supposed to be designed
as the flux of the momentum density ($\alpha^\sigma$ is the
quantum mechanical operator of the velocity, and the operator
$D_\mu$ has the kinetic momentum $P_\mu=m u_\mu$ as the prototype).
For now, we begin with a commonly used form
\begin{eqnarray}
T^{\sigma}_{~\mu}={i\over 2} \big(\psi^+ \alpha^\sigma
\overrightarrow{D_\mu} \psi - \psi^+ \overleftarrow{D^+_\mu}
\alpha^\sigma  \psi \big)~.
\label{eq:E3.6}\end{eqnarray}
which looks like a real quantity and is often believed to be
self-adjoint.  To avoid a premature discussion of the complicated
issue of self-adjointness that will arise later, I shall not dispute
this opinion, though some of the results will indicate
that, in general, this belief is wrong because the axial potential
in ${D_\mu}$ can be singular. Then it is not safe to have a
quantum mechanical operator that simultaneously acts in two adjoint
spaces  without taking care of a self-adjoint extension of this
(formally symmetric) operator. This will be done
in the next paper \cite{paper2},
where the condition of self-adjointness will be derived.

By virtue of the equations of motion, the following
identity holds
\begin{eqnarray}
\nabla_\sigma T^{\sigma}_{~\mu}=
i\psi^+ [D_\sigma, D_\mu ]\psi-
2mg \aleph_\mu \psi^+\rho_2 \psi \nonumber\\
+{i\over2}\nabla_\sigma \nabla_\mu(\psi^+\alpha^\sigma \psi)~.
\label{eq:E3.7}\end{eqnarray}
The last term in this equation is the result
of the following transformation of an anti-Hermitian form,\\
$\psi^+ \alpha^\sigma \overrightarrow{D_\mu}\psi +
\psi^+ \overleftarrow{D^+_\mu} \alpha^\sigma  \psi=
D_\mu(\psi^+ \alpha^\sigma\psi)= \nabla_\mu(\psi^+\alpha^\sigma\psi)$,
where a purely geometric relation
(\ref{eq:E2.31}), $~D_\mu\alpha_\sigma=0$, is employed and neither
the equations of motion nor a specific form of the spin connections
were used. The commutator in the brackets is the curvature tensor,
\begin{eqnarray}
[D_\sigma, D_\mu ]= D_\sigma D_\mu -D_\mu D_\sigma \nonumber\\
 =\partial_\mu\Gamma_\sigma-\partial_\sigma\Gamma_\mu
 +\Gamma_\sigma\Gamma_\mu-\Gamma_\mu\Gamma_\sigma .
\label{eq:E3.8}\end{eqnarray}
Since the spin connection $\Gamma_a$ is defined in terms of its tetrad
components, the primary objects are the
tetrad components of this tensor,
\begin{eqnarray}
D_{ab}=e^\alpha_aD_{\alpha\beta}e^\beta_b= D_aD_b-D_bD_a
+C^c_{ab}\Gamma_c~,
\label{eq:E3.9}\end{eqnarray}
where,
$$~D_aD_b-D_bD_a=\partial_b\Gamma_a-\partial_a\Gamma_b +
\Gamma_a\Gamma_b-\Gamma_b\Gamma_a $$
and the structure constants are
$C^c_{ab}=\omega^c_{~ab}-\omega^c_{~ba}$.
The commutator of covariant derivatives can be expressed in terms of
two gradient invariant tensors (the field strengths),
\begin{eqnarray}
 F_{\mu\nu}=\partial_\mu A_\nu -\partial_\nu A_\mu,~~~~
 {\cal U}_{\mu\nu}=\partial_\mu \aleph_\nu -\partial_\nu \aleph_\mu,
\label{eq:E3.10}\end{eqnarray}
and the Riemann curvature tensor. Since the vector current is conserved,
we have  $$\nabla_\sigma \nabla_\mu(\psi^+\alpha^\sigma \psi)
=R_{\sigma\mu}\psi^+\alpha^\sigma \psi,$$  where
$R_{\sigma\mu}=R^\lambda_{\cdot\sigma;\lambda\mu}$ is the Ricci tensor.
The final answer reads as follows,
\begin{eqnarray}
\nabla_\sigma T^{\sigma}_{~\mu}=
-eF_{\sigma\mu}[\psi^+\alpha^\sigma \psi]
-g{\cal U}_{\sigma\mu}[\psi^+\rho_3\alpha^\sigma \psi]\nonumber\\
-2gm\aleph_\mu~\psi^+\rho_2 \psi
+{i\over2}R_{\sigma\mu}(\psi^+\alpha^\sigma \psi)\nonumber\\
+{i\over 4}\psi^+\alpha_\sigma \sum_{c,d}
R^\sigma_{~\mu;cd}\rho_1\alpha^c\rho_1\alpha^d \psi~.
\label{eq:E3.11}\end{eqnarray}
The last two terms on the right hand side are imaginary and (if
written in natural units) proportional to Planck constant.
These terms exactly cancel each other.

So far, we have the divergence of the spinor energy-momentum tensor
on the left and the gradient-invariant Lorentz forces from the
electromagnetic field and a new axial field
on the right. The Lorentz force $~eF_{\sigma\mu}j^\mu~$
is due to the conserved probability current $j_a$ and it drives
the left and right currents in one common direction.  The axial
force, $~g{\cal U}_{\sigma\mu}{\cal J}^\mu$, is due to the not conserved
axial current ${\cal J}_a$, and it pulls left and right currents in
opposite directions. This new force must act between  any two Dirac
particles. Its effect is parity-odd; it has not been introduced
{\em ad hoc} and it cannot be rejected {\em a priori} within a
realm of various PNC atomic or nuclear phenomena. The extra term
which is proportional to the axial potential $\aleph_\mu$  is not
gradient invariant.

The next step is to convert this equation into the divergence of one
common energy momentum tensor for all fields in the system. Hence,
we need  the equations of motions for the fields $A_\mu$ and
$\aleph_\mu$.  Eqs.~(\ref{eq:E3.10}) immediately yield the first
(without the  sources) couple of  the Maxwell equations for the
field strengths $F_{\mu\nu}$ and ${\cal U}_{\mu\nu}$,
\begin{eqnarray}
\nabla_\lambda \epsilon^{\sigma\lambda\mu\nu}F_{\mu\nu}=0~,~~~~
\nabla_\lambda \epsilon^{\sigma\lambda\mu\nu}{\cal U}_{\mu\nu}=0~.
\label{eq:E3.12a}\end{eqnarray}

The equations that interconnect the fields and currents must be
{\em conjectured} from the existing data. We may rely on the
following facts: The hydrogen spectrum indicates that with great
accuracy the electron moves in the field with potential
$U=-e/r$ that satisfies the equation $\triangle U=-4\pi e\delta(r)$.
We can also refer to the data on PNC phenomena as an indication that the
axial current is involved in electron dynamics.

The potentials $A_\mu$ and $\aleph_\mu$ are Lorentz vectors. Therefore,
if we identify $~U=A_0~$ and $~-e\delta(r)=j_0~$ in the rest frame of the
atom, then the only possible Lorentz invariant form of the second
couple of the Maxwell equations is
\begin{eqnarray}
\nabla_\sigma F^{\sigma\mu}=ej^\mu=e[\psi^+\alpha^\mu \psi]~,
\label{eq:E3.12}\end{eqnarray}
so that $F_{\mu\nu}$ is the massless gradient-invariant Maxwell
field.  Actually, this choice is straightforward and simple because
the current $j_\mu$ is conserved. Then, we can present the
gradient-invariant Lorentz 4-force of the
electromagnetic field as the divergence of the energy-momentum tensor,
\begin{eqnarray}
e[\psi^+\alpha^\sigma \psi]F_{\sigma\mu}=\nabla_\lambda
\Theta^\lambda_{~~\mu}~,\nonumber\\
\Theta^\lambda_{~~\mu}=F^{\lambda\nu}F_{\nu\mu}+
(1/4)\delta^\lambda_\mu F^{\rho\nu}F_{\rho\nu}~.
\label{eq:E3.15}\end{eqnarray}
For the axial field $\aleph_\mu$, the choice of the second
equation is far less obvious because, except for the ``Lorentz
force'' $g{\cal U}_{\sigma\mu}{\cal J}^\sigma $, there is an
additional  not gradient-invariant term $2gm\aleph_\mu\psi^+\rho_2\psi
=g\aleph_\mu(\nabla_\nu {\cal J}^\nu)$.
The data on atomic spectra, which indicate that the {\it external}
field $\aleph_a$ does not depend on time and does not deviate from
the spherical symmetry, facilitates this choice. The only non-vanishing
components can be $\aleph_0(r)$ and $\aleph_r(r)$. Then,
${\cal U}_{0r}=-\partial_r\aleph_0(r)$ is the only non-zero component
of the field strength. This field must be strongly
confined near the nucleus, which requires a scale. A plausible form
is a Yukawa potential, so that
$$-\nabla^2\aleph_0(r) +M^2\aleph_0(r)=
g {\cal J}_0(r)~,$$
which, by the argument of Lorentz invariance, suggests that
$\aleph_\mu$ is a massive  neutral vector field,
\begin{eqnarray}
\nabla_\sigma {\cal U}^{\sigma\mu}+M^2\aleph^\mu=g {\cal J}^\mu=
g[\psi^+\rho_3\alpha^\mu \psi]~.
\label{eq:E3.13}\end{eqnarray}
(We have to put the mass term here to avoid
a conflict with the anti-symmetry of ${\cal U}_{\mu\nu}$.
Indeed, $\partial_\mu{\cal J}^\mu\neq 0$.)
Taking the covariant derivative of this equation we get
\begin{eqnarray}
M^2\nabla_\mu \aleph^\mu =
g \nabla_\mu {\cal J}^\mu =2~g~m{\cal P}~.
\label{eq:E3.14}\end{eqnarray}
Using (\ref{eq:E3.13}) and (\ref{eq:E3.5}) one can transform
the Lorentz force of the axial field in (\ref{eq:E3.11}) into
the divergence of its energy-momentum tensor,
\begin{eqnarray}
g[\psi^+\rho_3\alpha^\sigma \psi]{\cal U}_{\sigma\mu}
+2gm\aleph_\mu~\psi^+\rho_2 \psi
=\nabla_\lambda \theta^\lambda_{~\mu}~,
\label{eq:E3.16}\end{eqnarray}
where
\begin{eqnarray}
\theta^\lambda_{~\mu}={\cal U}^{\lambda\nu}{\cal U}_{\nu\mu}-
M^2\aleph^\lambda \aleph_\mu ~~~~~~~~~~~~\nonumber\\
+\delta^\lambda_\mu\bigg({1\over 4} {\cal U}^{\rho\nu}{\cal U}_{\rho\nu}+
{1\over 2}M^2\aleph^\rho \aleph_\rho\bigg)~.
\label{eq:E3.17}\end{eqnarray}
Putting Eqs.~(\ref{eq:E3.11})-(\ref{eq:E3.17}) together, one finds that
\begin{eqnarray}
\nabla_\lambda ( T^\lambda_{~~\mu} +\theta^\lambda_{~~\mu}
+\Theta^\lambda_{~~\mu})=0~,
\label{eq:E3.18}\end{eqnarray}
i.e., the total energy-momentum of three interacting fields, $\psi$, $A_\mu$,
and $\aleph_\mu$ is conserved, which is an additional indication that
the system of equations of motion is self-consistent.
In the next section we shall address the problem of the hydrogen atom and
find the effect of the static fields $\aleph_0(r)$ and $\aleph_r(r)$
on the solutions of the Dirac equation. Of particular interest will be
the quantum numbers of this problem and the existence of stationary
states.

\section{ Hydrogen with the Dirac proton \label{sec:Sec4}}
\renewcommand{\theequation}{4.\arabic{equation}}
\setcounter{equation}{0}

An obvious way to test the existence and properties of the new
axial field is to study its effect on atomic electrons treating the
field $\aleph_\mu(x)$ as a classical external field. We shall look
at the nucleus as a cluster of Dirac spinors. The electric and axial
currents of the nucleons are the sources of electric and
axial fields. The axial field, which has been introduced on
quite different premises,  very much resembles the neutral gauge field
of the standard model of electro-weak interactions. The neutron-rich
nuclei are known to have a larger weak charge. For the sake of
simplicity, the field of the nucleus will be taken as
time independent and spherically symmetric. The hydrogen atom is used
as an example where a new pattern naturally emerge in the course
of solving the Dirac equation with axial the field $\aleph$.

For hydrogen-like atoms we expect a perfectly spherically symmetric
geometry with a natural metric
$$ds^2=g_{\mu\nu}dx^\mu dx^\nu=
dt^2-r^2 d\theta^2-r^2\sin^2\theta d\varphi^2-dr^2~.$$
The tetrad vectors form a diagonal matrix,
and the only non-vanishing components of the Ricci rotation coefficients
are
\begin{eqnarray}
\omega_{122}=-{1\over r}~{\cos\theta\over\sin\theta}~,~~~
\omega_{232}={1\over r}~,~~~~~\omega_{311}=-{1\over r}.
\label{eq:E4.2}\end{eqnarray}
Assume that in addition to the Coulomb field $A_0(r)$, the nucleus is
also a source of a spherically symmetric axial field with the
temporal component $\aleph_0(r)$ and a radial one $\aleph_r(r)$.
These are the only components that can be present in a spherically
symmetric static object.
The Dirac equation is
\begin{eqnarray}
\big[i\partial_0 -eA_0 -g\rho_3\aleph_0 +g\sigma_3\aleph_r
-i\rho_3\sigma_3(\partial_r +{1\over r})~~~~~~~\nonumber\\
-i{\rho_3\sigma_1 \over r}(\partial_\theta+{1\over 2}\cot\theta)
-i{\rho_3\sigma_2\over r\sin\theta}\partial_\varphi
-m\rho_1 \big]\psi=0~.~~~
\label{eq:E4.3}\end{eqnarray}
In terms of a new unknown function,
$~\tilde{\psi}(r,\theta,\varphi)=r\sqrt{\sin\theta}\psi~$, this equation
becomes
\begin{eqnarray}
 \big[i\partial_0 -eA_0 -g(\rho_3\aleph_0-\sigma_3\aleph_r)
+\rho_3\sigma_3(-i\partial_r)~~~~~\nonumber\\
+{\rho_3 \over r}(-i\sigma_1\partial_\theta-i
{\sigma_2\over\sin\theta}\partial_\varphi) -m\rho_1 \big]\tilde{\psi}=0~.
\label{eq:E4.4}\end{eqnarray}
Equation (\ref{eq:E4.3}) is the Dirac equation in the tetrad basis. This
is a primary form of the spinor equation because it treats spinor field
according to its original definition as a Lorentz spinor
and a coordinate scalar.
(A conventional form, with spinors  related to Cartesian coordinates,
follows from this one, if at each space-time point the spinor is
Lorentz-rotated at a corresponding angle.) In order to find its
solution one has to separate the angular and radial variables in this
equation. This is known to be a somewhat tricky problem even in the
pure QED case (when $\aleph_a=0~$).

The Hermitian operators in Eq.~(\ref{eq:E4.4}) are  the tetrad components
of the momenta $p_3=-i\partial_r$, $p_1=-ir^{-1}\partial_\theta$
$p_2=-i(r\sin\theta)^{-1}\partial_\varphi$. The operators $p_1$ and
$p_2$ are clearly associated with the angular motion. If the coefficients
in this equation where not the matrices, it would have already been an
equation with the variables separated, which would match the perfect
spherical symmetry of the {\em external} fields $A_\mu$ and
$\aleph_\mu$. The problem is that the operators of radial and angular
momenta do not commute (they anti-commute,
$[\alpha_3p_3,(\alpha_1 p_1+\alpha_2 p_2)]_+=0~$).
A regular way to avoid this obstacle is as follows \cite{Weyl,Dirac1}.
One attempts to construct a minimal set of the operators that commute
with the Hamiltonian. For example, one can check that the commutator
$[\alpha_3p_3,\rho_1(\alpha_1 p_1+\alpha_2 p_2)]_-=0$, and take the
operator $\rho_1(\alpha_1 p_1+\alpha_2 p_2)$ as a generator of the
conserved quantum number. This trick works when $\aleph=0$ and it is
very instructive to see the details of its failure when $\aleph\neq 0$.

The conventional operator of the angular momentum is
$\vec{\cal L}=[\vec{r}\times\vec{p}]+\vec{\sigma}/2$. An additional
operator ${\cal L}=\vec{\sigma}\cdot\vec{\cal L}-1/2$ commutes
with the orbital momentum, $[{\cal L},(\vec{r}\times\vec{p})]=0$,
and has the properties, ${\cal L}({\cal L}-1)=[\vec{r}\times\vec{p}]^2$ and
${\cal L}^2=\vec{\cal L}^2+1/4$. Therefore, if $\kappa$ is an eigenvalue
of operator ${\cal L}$ we obviously
have $\kappa(\kappa-1)=l(l+1)$ and $\kappa^2>0$. On the other hand, if
${\cal L}'=\rho_1{\cal L}$, then $({\cal L}')^2={\cal L}^2$ and
these operators have the same sets of eigenvalues. In the tetrad
basis, these operators are
\begin{eqnarray}
{\cal L}= (-i\sigma_2\partial_\theta+
i {\sigma_1 \over\sin\theta}\partial_\varphi ),~~~
{\cal L}_3=-i \partial_\varphi +{1\over 2}\sigma_3~.
\label{eq:E4.7}\end{eqnarray}
In terms of the {\em auxiliary} operator ${\cal L}'$ (which
{\em has the same set of quantum numbers as the operator of the angular
momentum} but is a different operator) and the projection
${\cal L}_3$ of angular momentum, the Dirac equation becomes
\begin{eqnarray}
[i\partial_0 -eA_0 -g(\rho_3\aleph_0-\sigma_3\aleph_r)
+\rho_3\sigma_3(-i\partial_r) \nonumber\\
- \rho_2\sigma_3
{{\cal L}' \over r}-m\rho_1 ]\tilde{\psi}=0.
\label{eq:E4.5}\end{eqnarray}
If $\aleph_\mu=0$ then the operator ${\cal L}'$ commutes with the
Hamiltonian and we can require the wave function to be an
eigenfunction of the Hamiltonian and these two operators,
\begin{eqnarray}
{\cal L}'\tilde{\psi}=\kappa\tilde{\psi},~~~{\rm and}~~~
{\cal L}_3\tilde{\psi}=(m_z+1/2)\tilde{\psi}.
\label{eq:E4.6}\end{eqnarray}
This is a complete solution to the problem. The system
has two ``angular'' quantum numbers and a radial one.

Because the presence of the components $\aleph_0(r)$ and $\aleph_r(r)$
preserves  the spherical symmetry, we can try to look for the general
solution of the following form (the notation will become clear later on),
\begin{eqnarray}
\tilde{\xi}=\!\!\left( \begin{array}{c} f_{L\uparrow}(r,t)
{\cal Y}(\theta,\varphi) \\
 h_{L\downarrow}(r,t){\cal Z}(\theta,\varphi) \end{array}
                             \right)\!,~
\tilde{\eta}=\!\!\left( \begin{array}{c}  h_{R\uparrow}(r,t)
{\cal Y}(\theta,\varphi)\\
 f_{R\downarrow}(r,t){\cal Z}(\theta,\varphi)  \end{array} \right)\!.~
\label{eq:E4.8}\end{eqnarray}

 The properties of this form are easy to analyze after we substitute
 it into Eqs.~(\ref{eq:E4.5}) and (\ref{eq:E4.6})
 and put the results together
 (in each line the first equation ($=$) belongs to (\ref{eq:E4.5})
and the second equation ($==$) is taken from (\ref{eq:E4.6})),
\begin{widetext}
\begin{eqnarray}
([i\partial_0 -eA_0 -g(\aleph_0+\aleph_r)-i\partial_r]
f_{L\uparrow} {\cal Y}
= m h_{R\uparrow} {\cal Y} +{i \over r}(\partial_\theta-
{i \over\sin\theta}\partial_\varphi)h_{L\downarrow} {\cal Z}
==(m-i{\kappa \over r})h_{R\uparrow}{\cal Y}~,~~~(a)\nonumber\\
([i\partial_0 -eA_0 -g(\aleph_0-\aleph_r)+i\partial_r]
h_{L\downarrow} {\cal Z}
=m f_{R\downarrow} {\cal Z}+{i\over r}(\partial_\theta+
{i\over\sin\theta}\partial_\varphi)f_{L\uparrow} {\cal Y}==
(m+i{\kappa \over r})f_{R\downarrow} {\cal Z}~,~~~(b)\nonumber\\
([i\partial_0 -eA_0 +g(\aleph_0-\aleph_r)+i\partial_r]
h_{R\uparrow} {\cal Y}
= m f_{L\uparrow} {\cal Y}
-{i \over r}(\partial_\theta -{i \over\sin\theta}\partial_\varphi)
f_{R\downarrow} {\cal Z} ==
(m+i{\kappa \over r})f_{L\uparrow} {\cal Y}~,~~~(c)\nonumber\\
([i\partial_0 -eA_0 +g(\aleph_0+\aleph_r)-i\partial_r]
f_{R\downarrow} {\cal Z}
=m h_{L\downarrow} {\cal Z}- {i \over r}(\partial_\theta +
{i \over\sin\theta}\partial_\varphi)h_{R\uparrow} {\cal Y} ==
(m-i{\kappa \over r})h_{L\downarrow} {\cal Z}~,~~(d)
\label{eq:E4.11}\end{eqnarray}
\end{widetext}
One can easily see that these
equations do not go well together. Indeed, after the spinor
of the form (\ref{eq:E4.8}) is substituted into  Eqs.(\ref{eq:E4.6})
it becomes apparent that the angular variables can be
separated only when $f_{L\uparrow}=f_{R\downarrow}$
and $h_{R\uparrow}= h_{L\downarrow}$, which is not consistent with
the presence of the field $\aleph_a$ in Eqs.(\ref{eq:E4.5}).
In this case, the Dirac equation with the axial field breaks
up into two different (and thus, incompatible) systems of equations for
only two radial functions.

Nevertheless, just by inspection, one can verify that
the angular functions ${\cal Y}_{k,m}(\theta,\varphi)$ and
${\cal Z}_{k,m}(\theta,\varphi)$ that satisfy the  equations
\begin{eqnarray}
~~(\partial_\theta- {i \over\sin\theta
}\partial_\varphi){\cal Z}_{k,m}(\theta,\varphi)=
-k {\cal Y}_{k,m}(\theta,\varphi), \nonumber\\
(\partial_\theta+{i \over\sin\theta}\partial_\varphi)
{\cal Y}_{k,m}(\theta,\varphi)= k {\cal Z}_{k,m}(\theta,\varphi),~~
\label{eq:E4.11b}\end{eqnarray}
do separate angular variables in the Dirac equation (\ref{eq:E4.5}),
and do not separate them in Eq.~(\ref{eq:E4.6}).
The angular dependencies ${\cal Y}_{k,m}(\theta,\varphi)$ and
${\cal Z}_{k,m}(\theta,\varphi)$ are split between the components
of Dirac spinor that correspond to the opposite (outward and inward)
directions of the radial component of the electron spin, an eigen-state
of the operator $\sigma_3$ associated with the tetrad vector $e^3_a$.

After this separation of angular variables,  Eqs.~(\ref{eq:E4.11})
yield the following equations for the radial functions,
\begin{widetext}
\begin{eqnarray}
([i\partial_0 -eA_0 -g(\aleph_0+\aleph_r)-i\partial_r]
f_{L\uparrow}
=mh_{R\uparrow}-i{k \over r}h_{L\downarrow}~,~~~~~(a)\nonumber\\
([i\partial_0 -eA_0 -g(\aleph_0-\aleph_r)+i\partial_r]
h_{L\downarrow}
=mf_{R\downarrow}+i{k \over r}f_{L\uparrow}~,~~~~~(b)\nonumber\\
([i\partial_0 -eA_0 +g(\aleph_0-\aleph_r)+i\partial_r]
h_{R\uparrow}
=mf_{L\uparrow}+i{k \over r}f_{R\downarrow}~,~~~~~(c)\nonumber\\
([i\partial_0 -eA_0 +g(\aleph_0+\aleph_r)-i\partial_r]
f_{R\downarrow}
=mh_{L\downarrow}-i{k \over r}h_{R\uparrow}~.~~~~(d)
\label{eq:E4.11c}\end{eqnarray}
\end{widetext}
Now, it becomes clear that the axial field in spin connection causes
the qualitative change in the theory. Indeed, in the absence  of the
axial field $\aleph$, we have two identical couples of equations for
the radial functions that yield the well-known energy spectrum of the
hydrogen atom. Interestingly enough, this is exactly the spectrum
given by the Sommerfeld formula of the fine structure, i.e., the
solution of the relativistic Keppler problem with the Bohr-Einstein
quantization of of the radial motion and of the precession
of perihelion. ( The Bohr-Sommerfeld additional quantum number
$\kappa$ corresponds to the quantization of this
precession, and within this classical model there is no theoretical
arguments that would require us to reject the linear oscillatory orbit
with $\kappa=0$. It is excluded only because the spectral data
provide no evidence of stationary states with $\kappa=0$ in
hydrogen-like atoms.) Therefore, without the axial field,
the quantum eigenvalue problem yields a
classical relativistic answer, in accordance with the prototype
$u_\mu P^\mu=m$ of the Dirac equation. As long as the spinor
polarization is balanced between the left and
right components, there is not any visible spin effects and the
electron moves in an intermediate domain were the Maxwell field
absolutely dominates. The effects of the axial field are genuinely small
and proportional to the probability to find the electron inside the nucleus.
The mixing of levels that signals the possible parity non-conservation
is reasonably well described by perturbation theory in the
non-relativistic limit \cite{Khriplovich}.

In the presence of the axial field, the spinor polarization becomes more
agile and two new physical patterns develop. First, the angular
functions (\ref{eq:E4.11b}) are not connected {\em in any way} with the
angular momentum of the Dirac field \cite{Pauli}. Therefore, the angles
$\theta$ and $\varphi$
now belong to the  {\em internal space of parallel transport of the
spinor field along a closed spherical surface}. This, invisible from
the outside, internal space is very likely to have a non-Abelian symmetry
group and, possibly, a local gauge structure associated with it.
Therefore, the well-known non-Abelian theories may have a
realization within this scheme.

Second, since the angular functions ${\cal Y}(\theta,\varphi)$ and
${\cal Z}(\theta,\varphi)$ are not connected with the angular
momentum there is no reason to exclude the case $k=0$, which
indicates that a new physical pattern develops at short distances.
This choice corresponds to the physical situation when the Dirac
field is radially polarized in the internal geometry of a spherical
shell, so that the parallel transport of a Dirac spinor along a
sphere does not change it. A well-known prototype of such a state is
a fully occupied electron shell of a noble gas, which is symmetric to
the extent, that none of the electrons can be assigned individual
quantum numbers associated with the angular momentum. The only
topologically distinctive direction is the radial one. For such
states, the Dirac-type equation (\ref{eq:E4.11}) again splits into
two different independent systems for two polarization modes of the
Dirac wave function. One of these modes has only two non-zero
components, $\xi_1$ and $\eta_1$, which correspond to the positive
(outward) sign of the radial polarization. The second mode has only
$\xi_2$ and $\eta_2$ components. Its radial polarization is directed
towards the center (inward). Let us look for the stationary solutions
of these equations. For the outward mode we have
Eqs.~(\ref{eq:E4.11c}.{\it a,c}),
\begin{eqnarray}
~[E_{\uparrow} -eA_0 -g\aleph_0-g\aleph_r-i\partial_r] f_{L\uparrow}=
m h_{R\uparrow},~~(a)~~\nonumber\\
~[E_{\uparrow} -eA_0 +g\aleph_0-g\aleph_r+i\partial_r]h_{R\uparrow}=
m f_{L\uparrow},~~(c)~~
\label{eq:E4.12}\end{eqnarray}
For the inward mode (Eqs.~(\ref{eq:E4.11c}.{\it b,d}) )
we have the same system with the opposite signs
of $\aleph_0$ and $\aleph_r$.

The wave
functions of the inward and outward polarization modes have different
time dependences. It is instructive to see by how much these new
solutions deviate from a purely Coulomb problem. The equation for the
static component of the axial potential produced by a compact spinor
configuration will be derived in the next paper\cite{paper2}, and it
will yield the expression (\ref{eq:E4.19}) below. At the moment, we
shall take a short cut by noticing that in a perfectly spherical
static case the field $\aleph$ can have only two components
$\aleph_0(r)$ and $\aleph_r(r)$, so that only ${\cal U}_{0r}(r)\neq
0$. Then by virtue of (\ref{eq:E3.13}) and (\ref{eq:E3.14}) we have a
short-range Yukawa field $\aleph_0(r)$ with the source $g {\cal
J}_0(r)$ which is spread at a distance $r_w\sim 1/M$, and a radial
field $\aleph_r(r)$ for $r>r_{max}$,
\begin{eqnarray}
g\aleph_r(r)= - {Q(r)\over r^2}= -{1\over r^2}{2 g^2 m \over
M^2} \int_0^{r_{max}}{\cal P}(r)r^2~dr.~~
\label{eq:E4.19}\end{eqnarray} The latter is determined from ``Gauss'
law'' (\ref{eq:E3.14}) for the field $\overrightarrow{\aleph}$ with
the nuclear pseudoscalar density as a source. Depending on how
singular the distribution of this density is, the shape of the potential
$\aleph_r$ can range from nearly homogeneous within the nucleus to as
singular as $r^{-2}$.

The position of the potential $\aleph_0$ in Eqs.~(\ref{eq:E4.12})
is such that one can absorb the $\aleph_0(r)$ into
a phase factor by means of the substitution,
\begin{eqnarray}
\{f,h\}_{\uparrow\downarrow}=\{\tilde{f},\tilde{h}\}_{\uparrow\downarrow}
e^{\pm ig\int_0^r \aleph_0(r)dr}~.
\label{eq:E4.14}\end{eqnarray}
Taking an attractive Coulomb potential with
the point-like charge $~+Ze~$ for $~A_0(r)$ and $~\aleph_r(r)~$ from
Eq.~(\ref{eq:E4.19}) we obtain two systems of equations for two
topologically distinct modes,
\begin{eqnarray}
\bigg(i\partial_0 -{Ze^2\over r}\mp { Q(r)\over r^2}
-i\partial_r \bigg)\tilde{f}_{\uparrow\downarrow}=
m \tilde{h}_{\uparrow\downarrow}~,\nonumber\\
\bigg(i\partial_0 -{Ze^2\over r}\mp { Q(r)\over r^2}
+i\partial_r \bigg)\tilde{h}_{\uparrow\downarrow} =
m \tilde{f}_{\uparrow\downarrow} ~.
\label{eq:E4.17}\end{eqnarray}
 Let us assume that $Q(r)>0$
within some range of radius $r$ and take a close look at these
equations. The outward $\uparrow$-component of the Dirac spinor
is universally attracted and the inward $\downarrow$-component
is universally repulsed from
the positive central pseudoscalar charge. From this perspective,
the distribution of the pseudoscalar density in the integrand of the
"charge" $Q$ is the most intriguing issue. If the
pseudoscalar density is localized in a very small volume, the
potential $ \pm Q/r^2$ can overwhelm the Coulomb potential.
It will be strongly repulsive for the $\downarrow$-component and strongly
attractive for the $\uparrow$-component, even reaching the critical
boundary of falling onto the center. One may wonder if these
dynamics can lead to the formation of stable and/or metastable
states, and what the quantum numbers of these states are and how rich the
spectrum of these states is.  The radial dependence of
$g\aleph_r(r)$ is such that, unless it is screened at very short
distances, this potential always wins. While the Coulomb potential can
become catastrophically attractive only at $Z\geq 137$, when it
overwhelms the centrifugal barrier, nothing can withstand the
attractive $-Q/r^2$ at small $r$, regardless of the magnitude of
$Q$. ( V. Gribov made an attempt to incorporate this phenomenon into his
picture of quark/color confinement\cite{Gribov}.) In this case
of a singular potential,  the
conventional wisdom regarding what is attracted or repelled
 does not work any more. For example, a  positive
singular potential can be attractive for states of negative
energy with no lower limited to the energy spectrum. Then the
Hamiltonian of the Dirac equation is not self-adjoint. This
observation will be the starting point of the subsequent paper
\cite{paper2}, where the approach initiated here will be reconsidered
once again with special attention to the self-adjointness of the Dirac
operators and to the {\em existence of meaningful dynamics} in the
presence of so singular potentials.

\section{ Non-relativistic limit \label{sec:Sec5}}
\renewcommand{\theequation}{5.\arabic{equation}}
\setcounter{equation}{0}

At the atomic scale, the relativistic effects are small corrections
unless polarization (spin) effects are involved.  A formal $v/c$
expansion of the Dirac equation naturally leads to the modified form
of the Pauli equation. Let us proceed from the Dirac equation with
the axial field
(\ref{eq:E3.2}) in Cartesian coordinates, where the Ricci coefficients
are absent,
\begin{eqnarray}
\bigg(i\partial_0 -{e\over\hbar c}A_0-{g\over\hbar c}\rho_3
\aleph_0\bigg)\psi ~~~~~~~~~~~~~~~~~~~\nonumber\\
+\vec{\alpha}\bigg(i\vec{\nabla}-{e\over\hbar c}\vec{A}-
{g\over\hbar c}\rho_3 \overrightarrow{\aleph}\bigg)\psi=
{m c\over\hbar}\rho_1 \psi~.
\label{eq:E5.1}\end{eqnarray}
In terms of the large and small spinor components,
$$\phi= e^{-imc^2t/\hbar}(\xi+\eta),~~{\rm and}~~
\chi= e^{-imc^2t/\hbar}(\xi-\eta)~,$$
(i.e., in standard representation) these equations become
\begin{eqnarray}\label{eq:E5.3}
[i\hbar\partial_t -eA_0-g
(\vec{\sigma}\cdot\overrightarrow{\aleph})]\phi
~~~~~~~~~~~~~~~~\nonumber\\
=[\vec{\sigma}\cdot(c~\vec{p}+ e\vec{A})+g \aleph_0]\chi~,\\
\big[i\hbar\partial_t -eA_0-g (\vec{\sigma} \cdot \overrightarrow{\aleph})
+2m_0 c^2\big]\chi ~~~~~~~~~~\nonumber\\
=[\vec{\sigma}\cdot(c~\vec{p}+ e\vec{A})
+g \aleph_0]\phi,~~
\label{eq:E5.4}\end{eqnarray}
where $\vec{p}=-i\hbar\vec{\nabla}$. In the lowest approximation with
respect to $1/c$,  equation (\ref{eq:E5.4})yields
$$\chi ={1\over 2mc}[\vec{\sigma}\cdot(\vec{p}+
{e\over c}\vec{A})+g \aleph_0]\phi, $$
and inserting this approximation into
Eq.~(\ref{eq:E5.3}) we arrive at the Pauli type equation,
\begin{eqnarray}
i\hbar{\partial\phi\over \partial t}=eA_0\phi+{1\over 2m}
(\vec{p}+ {e\over c}\vec{A})^2\phi -
{e\hbar\over 2m c}(\vec{\sigma} \cdot \vec{B})\phi \nonumber\\
+g(\vec{\sigma} \cdot \overrightarrow{\aleph})\phi
-{ig\hbar\over 2mc}[(\vec{\sigma} \cdot \vec{\nabla})\aleph_0
+\aleph_0(\vec{\sigma} \cdot \vec{\nabla})]\phi \nonumber\\
-{eg\over mc^2}\aleph_0(\vec{\sigma} \cdot \vec{A})\phi
+{g^2\over2mc^2}\aleph_0^2 \phi~.
\label{eq:E5.5}\end{eqnarray}
The first three terms on the right-hand side are the well known terms
of the Pauli equation. Despite the presence of the axial field, the kinetic
Schr\"{o}dinger type term captures only the convection pattern
and it is not altered by the axial field.
The fourth and the fifth terms are due to the vector and scalar parts
of the axial potential, respectively. The latter is frequently used in
non-relativistic calculations. It is attributed to the interaction of
axial electron current with the vector part of the electro-weak current
and gives a major
contribution to the PNC atomic phenomena in heavy atoms. The term
$g(\vec{\sigma} \cdot \overrightarrow{\aleph})$
 corresponds to the interaction between
the electron and nuclear spin. A detail analysis of spin polarization
effects that lead to PNC  in atomic physics can be found in the book
\cite{Khriplovich}. If $\overrightarrow{\aleph}$ is considered in the
wider context of an arbitrary external field then one can introduce
an effective, with respect to its action on atomic systems, magnetic
equivalent of this field,
\begin{eqnarray}
 \vec{B}_{\aleph}={2mc\over\hbar}~{g\over e}~\overrightarrow{\aleph}
={g\over \mu_{ B}}~\overrightarrow{\aleph}~.
\label{eq:E5.6}\end{eqnarray}
Its effect is qualitatively indistinguishable from a true
magnetic field.

\end{document}